\documentclass[preprint,prl,showkeys,showpacs]{revtex4}
\newcommand{\beq}{\begin{equation}}
\newcommand{\eeq}{\end{equation}}
\usepackage{graphicx}
\usepackage{graphics}
\usepackage{latexsym}
\usepackage{subfigure}
\usepackage{multirow}
\usepackage{dcolumn}
\begin{document}

\title{A cellular automaton identification of the universality classes of spatiotemporal intermittency}

\author{Zahera Jabeen}
\email{zahera@physics.iitm.ac.in}
\author{Neelima Gupte}
\email{gupte@physics.iitm.ac.in}
\affiliation{Indian Institute of Technology-Madras, Chennai, India}
\date{\today}

\begin{abstract}

The phase diagram of the coupled sine circle map lattice shows
spatio-temporal intermittency of two distinct types: spatio-temporal 
intermittency of the directed percolation (DP) class, and spatial
intermittency which does not belong to this class. 
These two types of behaviour are seen to be special cases of the
spreading and non-spreading regimes seen in the system. In the spreading regime, each site can infect its neighbours permitting an initial disturbance to spread, whereas  in the non-spreading regime no infection is possible. 
The two regimes are separated by a line which we call the  infection line. 
The coupled map lattice can be mapped on to an equivalent cellular automaton which shows a  transition from a probabilistic cellular automaton (PCA) to a deterministic cellular automaton (DCA) at the infection      
line. Thus the           
existence of the DP and non-DP universality classes in the same system      
is  signalled by the PCA to DCA transition. We also discuss the
dynamic origin of this transition.  

\end{abstract}  
\maketitle

The identification of the universality class of spatiotemporal
intermittency \cite{daviaudpirat} in spatially extended systems has been a long standing
problem in the literature. Early conjectures argued that the transition 
to spatio-temporal intermittency is a second order phase transition, and
the transition falls in the same universality class as directed
percolation \cite{Pomeau}. 
This conjecture has
become the central issue in a long-standing debate
\cite{Rolf,Chate,Grassberger}, which is still not
completely resolved.
  
Studies of the coupled sine circle map lattice have 
thrown up a number  of intriguing observations of relevance to this 
problem \cite{JanZ1Z2}. 
This system has regimes of spatio-temporal intermittency (STI) with critical 
exponents which fall in the same universality class as directed
percolation (DP), as well as regimes of spatial intermittency (SI) which do not
belong to the DP class. Both these regimes lie on the bifurcation
boundaries of the synchronised solutions of the map. The
spatio-temporally intermittent regime seen here has an absorbing laminar
state, i.e. a laminar site remains
laminar unless infected by a neighbouring turbulent site. The burst
states spread and can percolate through the entire lattice. The system
shows a convincing set of directed percolation exponents in this regime \cite{JanZ1Z2}.   
In the spatially intermittent regime, the laminar sites are frozen in
time and the burst sites show temporally periodic or quasi-periodic
behaviour. The laminar sites do not get infected by neighbouring
turbulent sites.
Hence, the spatially intermittent state is non-spreading and does not 
show directed percolation exponents. Thus, both DP and non-DP behaviour can be seen for
different parameter regimes of the same system.     

In the present paper, we show that the infective directed percolation behaviour of
spatio-temporal intermittency and the non-infective behaviour of spatial 
intermittency are special cases of the more general spreading to
non-spreading transition seen in this system. The spreading and
non-spreading regimes are separated by a line which we call the
infection line. Above the infection line, the burst states can infect
neighbouring laminar states and spread through the lattice, whereas 
below this line the burst states cannot infect their neighbours and the
non-spreading regime is seen. The infection line intersects the
bifurcation boundary of the synchronised solutions. 
Intermittent
solutions are seen along this boundary, with the DP type of STI being seen above the
infection  line, and the non-DP SI being seen below the infection
line. Spreading and non-spreading solutions are also seen off the
bifurcation boundary. However, the distribution of laminar lengths shows power-law scaling 
only for parameter values which are very close to the bifurcation
boundary, and falls off exponentially as the 
parameter values get more distant from the bifurcation boundary. 
Other exponents associated with DP behaviour are also observed only along the
bifurcation boundaries.
Further insights into the spreading to non-spreading transition are
obtained by mapping the coupled map lattice (CML) onto a cellular automaton. The spreading to
non-spreading transition seen across the infection line maps
on to a transition from a probabilistic cellular automaton to a
deterministic cellular automaton. Thus the change from 
spreading to non-spreading behaviour seen in the CML is reflected in 
this transition. We also provide a pointer to the dynamic origin of this 
transition. 
\begin{figure}[!t]
\includegraphics[width=10.5cm,height=7.cm]{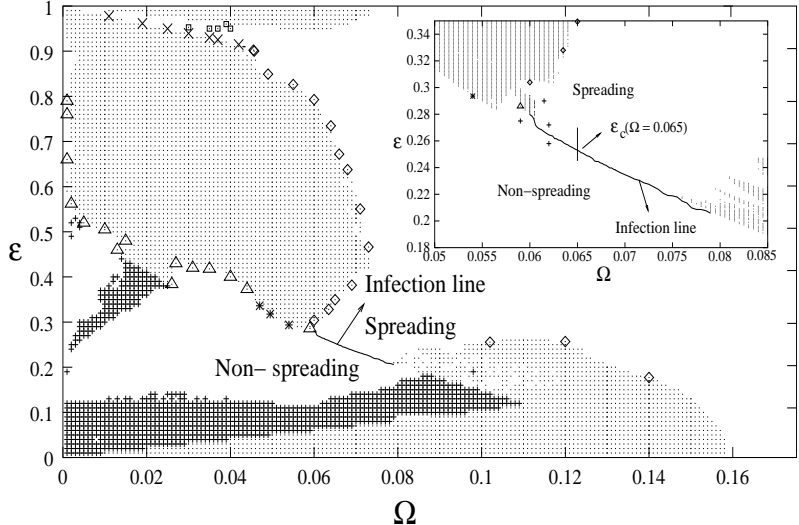}
\caption{ 
shows the phase diagram of the CML.  Spatiotemporal intermittency  of the
DP class is
seen at  the $\Diamond$-s. Spatial intermittency with quasi-periodic
bursts is seen at the
$\triangle$-s,  SI with periodic bursts  at  the $\times$-s and$\ast$;
STI with travelling wave laminar states and solitons at the boxes
($\Box$).
The spreading and the
non-spreading regimes  separated by the infection line are shown.
The 
region of cross-over from STI of the DP class 
(indicated by $\Diamond$-s), to SI with quasi-periodic bursts 
(indicated by $\triangle$-s) is shown in the inset. 
 \label{pdsns}}
\end{figure}

The coupled sine circle map lattice studied here is known to model the
mode-locking behaviour \cite{gauri2} seen in coupled
oscillators, Josephson Junction arrays etc. 
The model is defined by
the evolution equations
\beq
x_i^{t+1}=(1-\epsilon)f(x_i^t)+\frac{\epsilon}{2}[ f(x_{i-1}^t) +
f(x_{i+1}^t) ]\pmod{1}
\label{evol}
\eeq

where the index $i$ is a discrete site index which runs on a one dimensional lattice of $N$ sites, and $t$ s a discrete time index. The parameter 
$\epsilon$ is the strength of the coupling between the site $i$ and
its two nearest neighbours. The local on-site map, $f(x)$ is the sine
circle map defined as
$f(x)=x+\Omega-\frac{K}{2\pi}\sin(2\pi x)$,
where, $K$ is the strength of the nonlinearity and $\Omega$ is the
winding number of the  single sine circle map in the absence of the
nonlinearity. We study the system with periodic boundary conditions in
the parameter regime  $0 \leq \Omega \leq \frac{1}{2\pi}$ (where
the single circle map has temporal period
1 solutions), $0 \leq \epsilon \leq 1$ and $K=1.0$.
The phase diagram of this model evolved with random initial conditions 
is shown in Fig. \ref{pdsns}.
\begin{figure}[!t]
\begin{tabular}{cc}
{\hspace{-.1in}\includegraphics[width=5.5cm,height=5.cm]{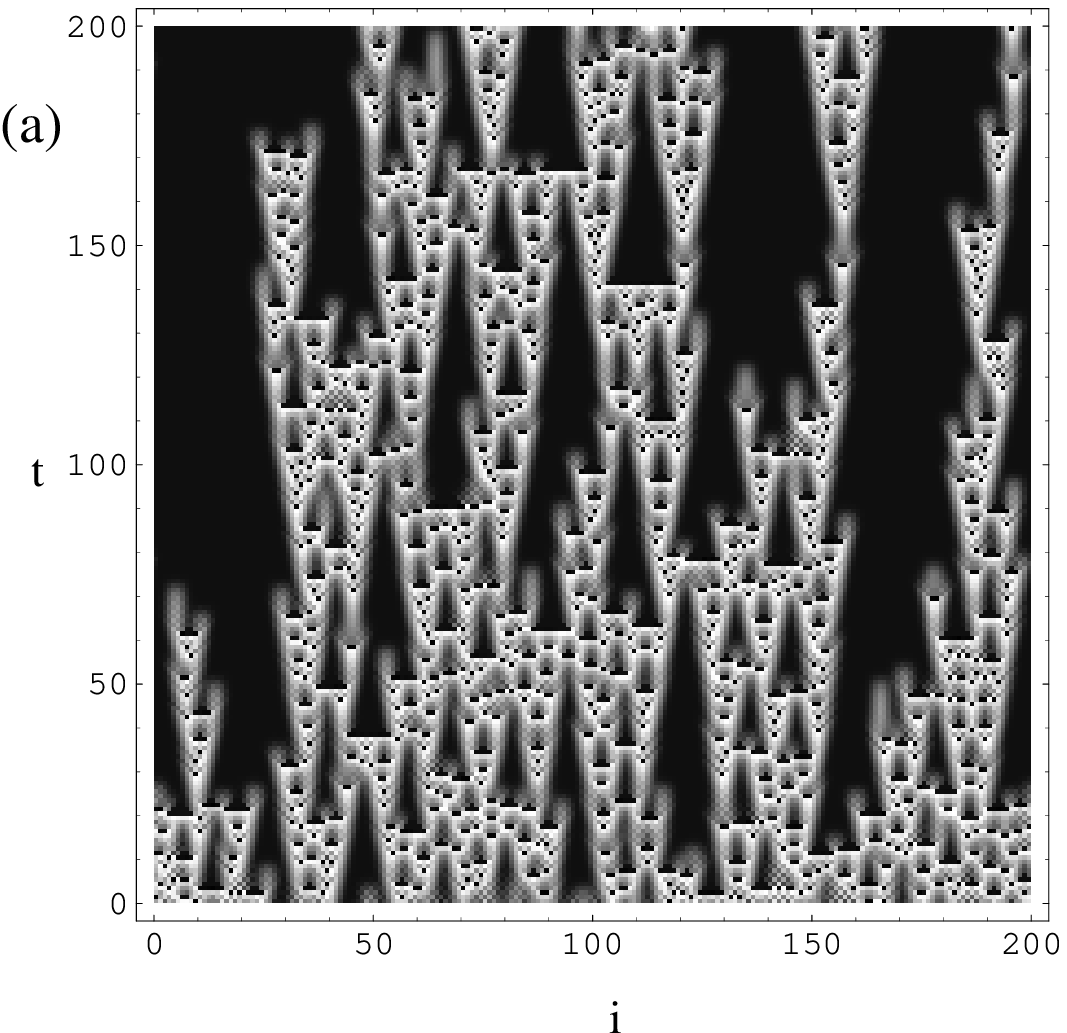}
}&
{\hspace{-.0cm}\includegraphics[width=5.5cm,height=5.cm]{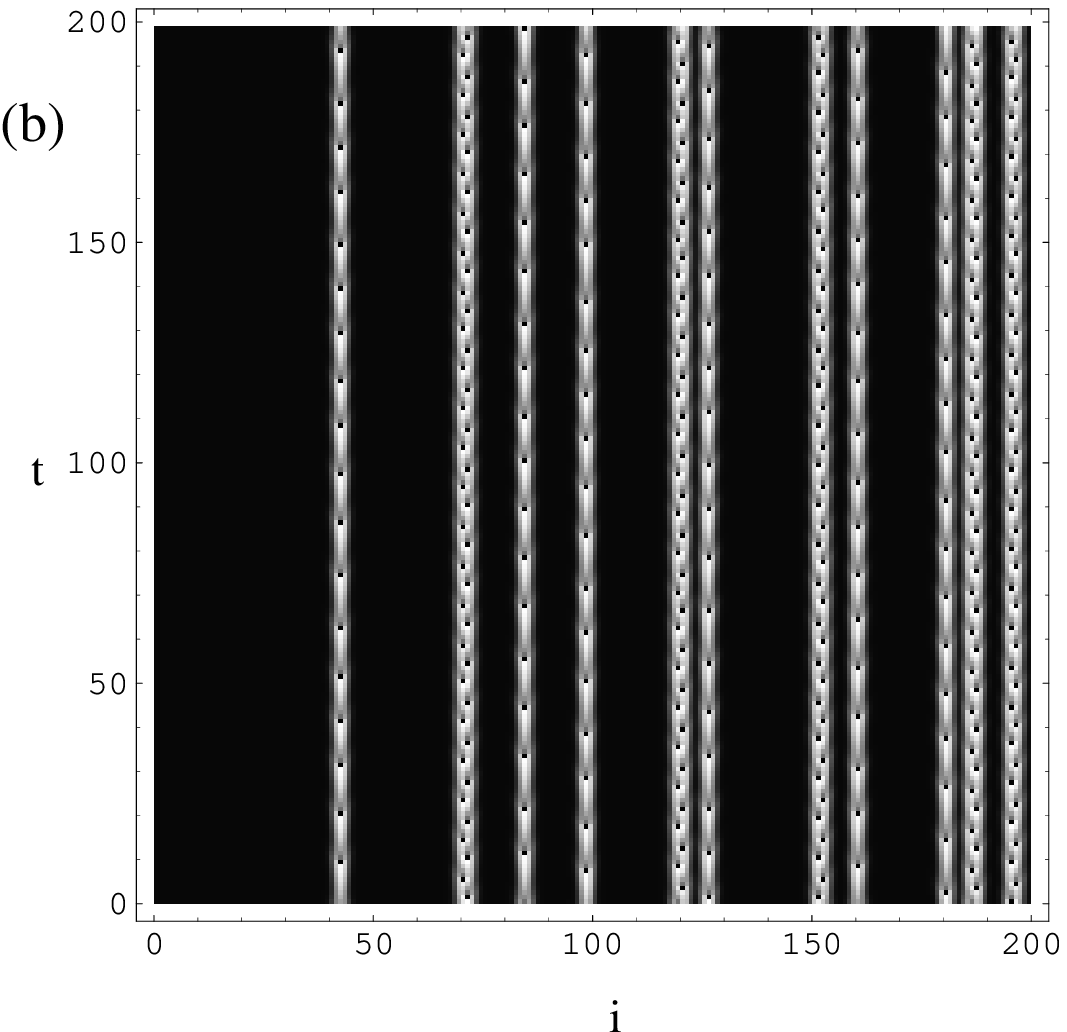}
}
\\
\end{tabular}
\vspace{-.2in}
\caption{ The space time plots of (a) STI with synchronized laminar state and turbulent bursts seen at $\Omega=0.06, \epsilon=0.7928$.(b) SI with synchronized laminar state and quasi-periodic bursts seen at $\Omega=0.031, \epsilon=0.42$. 
\label{stplot}} 
\end{figure}

\begin{figure}[!b]
\begin{tabular}{cc}
{\hspace{-.0in}\includegraphics[width=5.5cm,height=5.cm]{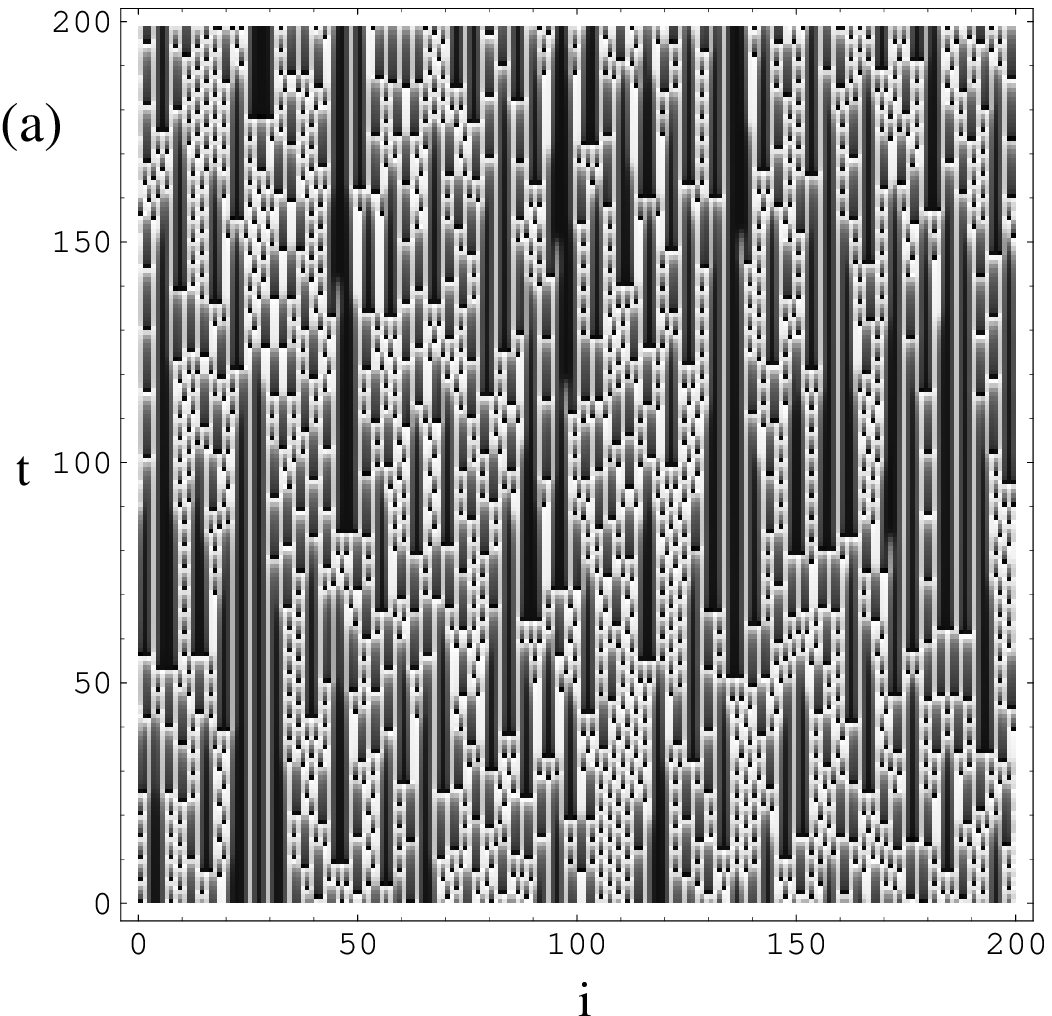}
}&
{\hspace{-.0cm}\includegraphics[width=5.5cm,height=5.cm]{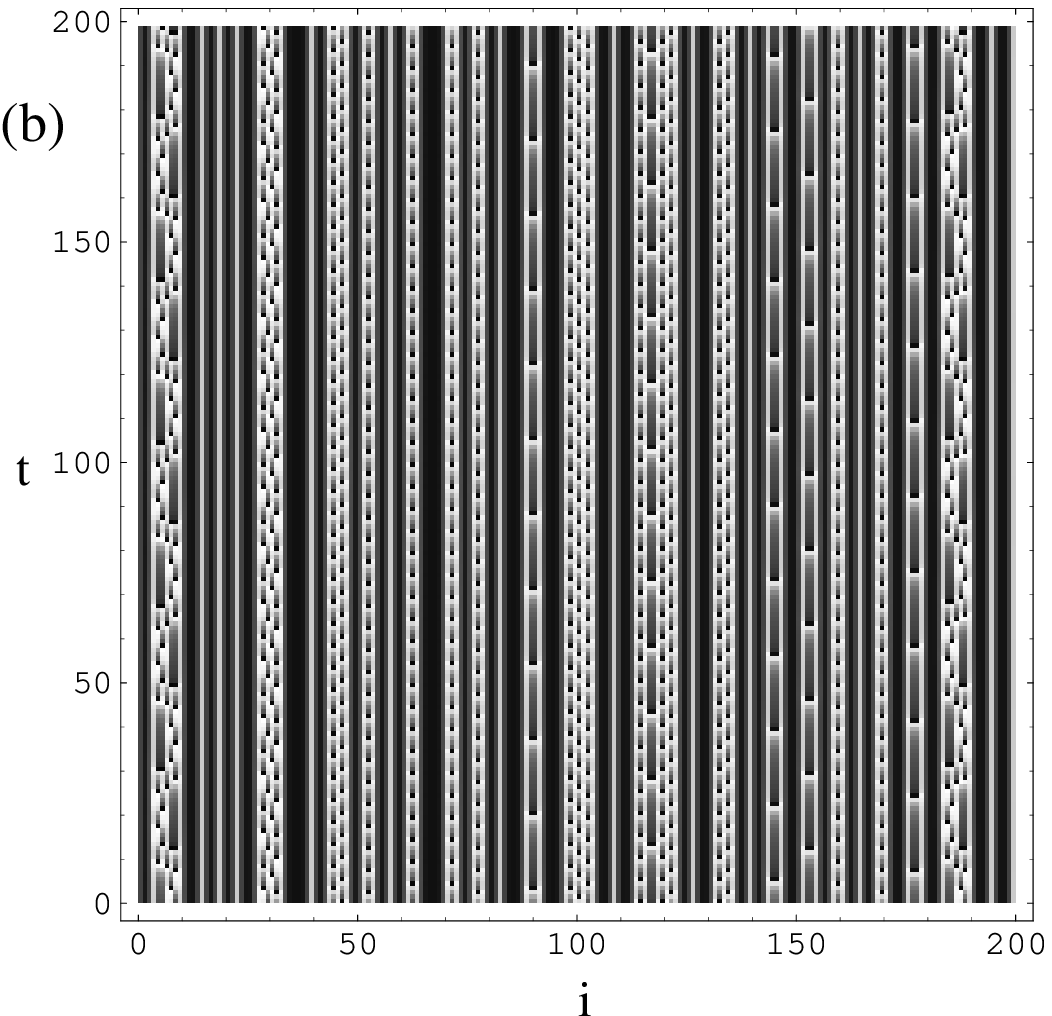}
}
\\
\end{tabular}
\vspace{-.2in}
\caption{ The space time plots seen (a) above the infection line at $\Omega=0.062, \epsilon=0.272$ and  (b) below the infection line at $\Omega=0.062, \epsilon=0.258$.  \label{stsns}} 
\end{figure}

The synchronised fixed point solutions
$x^{\star}=\frac{1}{2\pi}\sin^{-1}(\frac{2\pi\Omega}{K})$ are seen in
the regions indicated by dots in the phase diagram.
The infection line  seen in the figure separates
the remaining part into the 
spreading and  non-spreading regions.
The space-time plots of the solutions seen in both these regions
show co-existing laminar states and turbulent states (See Fig. \ref{stplot},\ref{stsns}). 

\begin{figure}[!t]
\begin{tabular}{cc}
\includegraphics[width=5.5cm,height=5.cm]{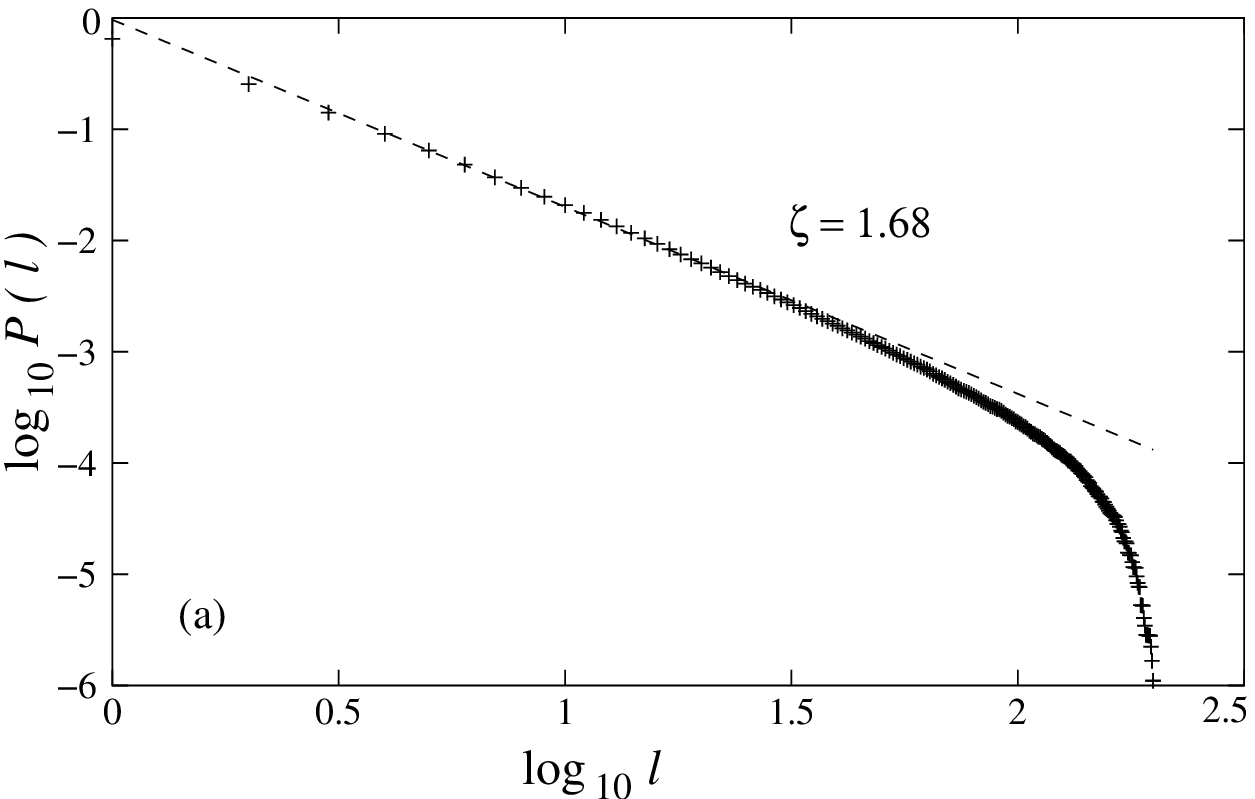}&
~~~\includegraphics[width=5.5cm,height=5.cm]{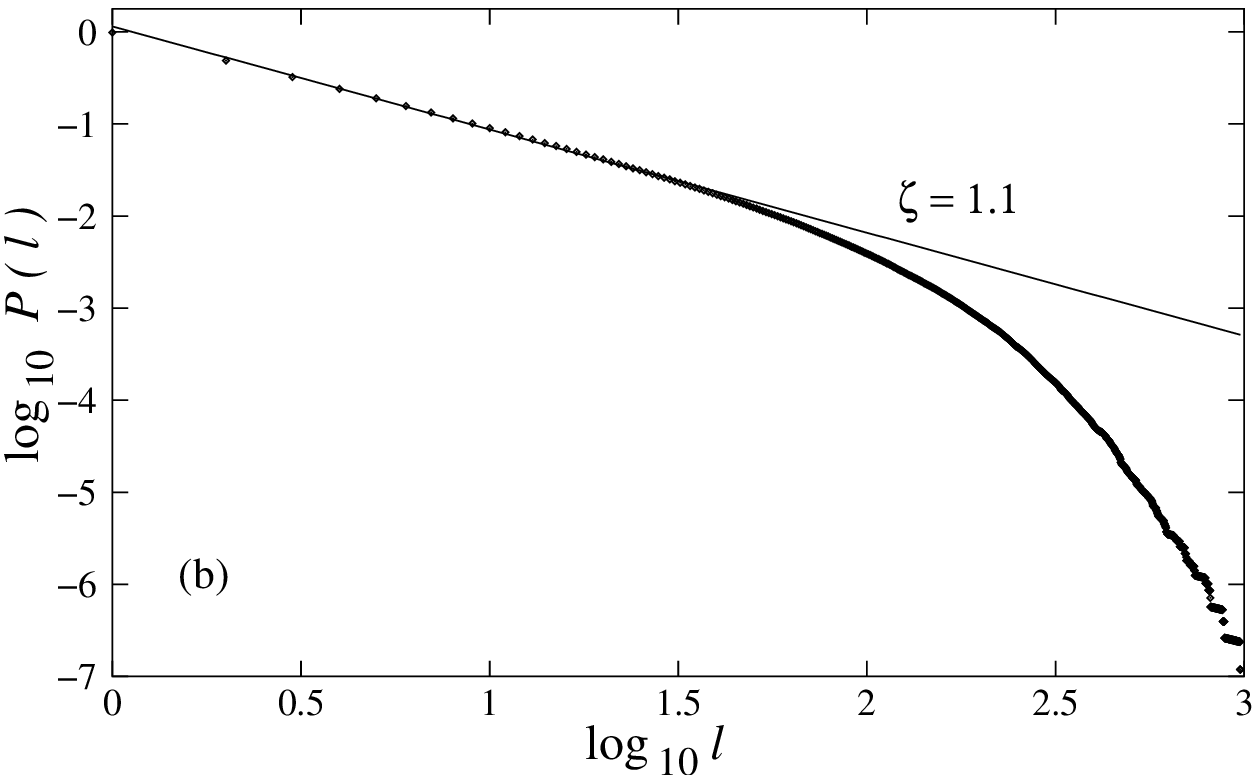}\\
\end{tabular}
\caption{ shows the laminar length distribution of (a) STI of the DP class seen at $\Omega=0.06, \epsilon=0.7928$ and (b) SI with quasi-periodic bursts at $\Omega=0.04, \epsilon=0.402$. The data have been obtained for a $10^4$ site lattice and are averaged over $50$ initial conditions.  \label{lldns}}
\end{figure}

The power law behaviour of the distribution of laminar lengths seen at
the DP points (indicated by
diamonds on the bifurcation boundary above the infection line)
is shown in  Fig. \ref{lldns}(a).
This scales   with an exponent
$\zeta=1.67$, characteristic of directed
percolation. A full set of directed percolation exponents is seen at
these points \cite{JanZ1Z2}. The exponential fall off of the laminar length
distributions for the   
spreading solutions  off 
the bifurcation boundary,  
is shown in
Fig. \ref{lldns}(b). Below the infection line, in the non-spreading
regime,
the laminar sites are 
the synchronised fixed point $x^*$ and the burst sites can be temporally 
frozen, periodic or aperiodic. 
The power-law scaling of the laminar length distribution for  spatial         
intermittency is also shown in Fig. \ref{lldns} (a). Here the 
distribution scales with
exponent $\zeta=1.1$, distinct from the DP value. The exponential
decrease of 
this distribution seen for other non-spreading
solutions at points off
the bifurcation boundary is shown in  Fig. \ref{lldns}(b). Thus the sine circle map CML shows a  transition from a spreading regime
to a non-spreading regime. In order to gain further insights into this
transition,  
we  map the CML to a stochastic model,
a probabilistic cellular automaton of the Domany-Kinzel type \cite{Chate,domanykinzel}.\\

The equivalent cellular automaton, defined on a one dimensional lattice
of size $N$, is set up to mimic the dynamics of the laminar and burst
states.   
The state variable $v_i^t$ at site
$i$ and at time $t$ takes values
$v_i^t=0$ if the site is in the laminar state, and $v_i^t=1$ 
if the site is in the burst state. 
By the CML 
evolution equation  (\ref{evol}), the state of the variable  
at site $i$ at time $t+1$ depends on the state of the variables
at sites $i$, $i-1$ and $i+1$ at time $t$.  Hence, the
probability of the site $i$ at time $t+1$ being in the burst state
depends on the state of the sites $i-1,i$ and $i+1$ at time $t$. 
We therefore define the CA dynamics in this system
by the conditional probability $P(v_i^{t+1}|v_{i-1}^t,v_i^t,v_{i+1}^t)$.
There are $2^3$ possible configurations and 
the symmetry between
the sites $i-1$ and $i+1$ in the CML equation is used to  obtain the  effective probabilities $p_k$'s, and define the  CA rules as  $p_0=P(1|000)$ , $p_1=P(1|001)=P(1|100)$, $p_2=P(1|010)$, $p_3=P(1|011)=P(1|110)$, $p_4=P(1|101)$ and $p_5=P(1|111)$

The direct connection between equivalent CA and the CML of eq. (\ref{evol}) is
set up by estimating the probabilities $p_k$ from the numerical evolution 
of the CML from random initial conditions from a uniform distribution
over the $(0,1)$ interval for a given set of parameter values. 
The probabilities are estimated by finding  
the fraction of sites $i$ which are in the burst state $v_i^{t+1}=1$ at
at time $t+1$, given
that the site $i$ and its nearest neighbours $i-1$ and
$i+1$ existed in state $k$ at time $t$. 
That is, the probability $p_k$ is estimated using
$p_k=\frac{N_1^{k}}{N_0^k+N_1^k}$, where
 $N_0^k$ and $N_1^k$ are the number of
sites  which at time $t+1$ exist in the laminar states
($v_i^{t+1}=0$) and the burst states ($v_i^{t+1}=1$) respectively and at
time $t$ were the central sites of the configuration $k$.
These probabilities, were extracted from a  CML of size $N=2000$
averaged over $20000$ timesteps discarding  a transient of $30000$
timesteps. The probabilities $p_k$ in the spreading and
non-spreading regimes in the phase diagram are listed in the
Table \ref{pkdpsi}.
\begin{table}[!b]
\begin{tabular}{ccccccccc}
\hline
&$\Omega$&$\epsilon$ & $p_0$ & $p_1$ & $p_2$ & $p_3$ & $p_4$ & $p_5$\\
\hline
\multirow{2}{*}{S(DP)}&~0.060~ & ~0.7928~ &0.0 &~0.220~ &0.0 &~0.933~ &~0.627~ &~0.984~\\
&~0.073~ &~0.4664~ &0.0 &~0.150~ &0.0 &~0.938~ &~0.439~ &~0.993~\\
\hline

\multirow{2}{*}{S}&~\multirow{2}{*}{0.070}~&~0.264~ &  0.0 &  ~0.140~ &  0.0 &  ~0.982~ &  ~0.391~ &  ~0.999~ \\
&&~0.248~ &  0.0 &  ~0.050~ &  0.0 &  ~0.989~ &  ~0.160~ &  ~0.999~ \\
\hline
\multirow{2}{*}{NS}&~\multirow{2}{*}{0.070}~& ~0.232~ &  0.0 &  ~0.000~ & 0.0 &  ~1.000~ &  ~0.000~ & ~1.000~ \\
&& ~0.228~ &  0.0 &  ~0.000~ & 0.0 &  ~1.000~ &  ~0.000~ & ~1.000~ \\
\hline
\multirow{2}{*}{NS(SI)}&~0.031~& ~0.420~ & 0.0 & ~0.000~& 0.0 &1.000~&~0.000~&~1.000~\\
& ~0.044~ & ~0.373~ & 0.0 & ~0.000~& 0.0 &1.000~&~0.000~&~1.000~\\
\hline

\end{tabular}
\caption{ shows the probabilities $p_k$'s obtained in the spreading (S), non-spreading (NS) regimes and at DP and SI points averaged over $50$ initial conditions.\label{pkdpsi}}
\end{table}
It is clear that that the probability $p_0=P(1|000)$ is equal to zero in both regimes. This
is the condition for an absorbing state where 
a laminar site with two laminar neighbours cannot
spontaneously evolve into a burst state.  
We also see that $p_2=P(1|010)=0$ so that a burst site with two laminar
neighbours always goes into a laminar state, i.e. the laminar neighbours
suppress the central burst site. The probabilities
$p_1=P(1|001)=P(1|100)$ and  $p_4=P(1|101)$ are essentially infection
probabilites by which a  laminar site is infected by its burst neighbour
or neighbours to change to a burst site.

It is clear from Table \ref{pkdpsi} that these probabilities show
drastically different behaviour in the spreading and non-spreading
regimes. 
In the case of the spatio-temporal intermittency with directed percolation exponent, which lies in the spreading regime of the
phase diagram, these infection probabilities $p_1$ and $p_4$ lie 
in the open interval $(0,1)$. Therefore, in the spreading regime, the dynamics is described by 
 a probabilistic cellular automaton wherein the CA rules are
probabilistic in nature.

In the case of spatial intermittency which lies in the non-spreading regime, 
the probabilities obtained take the values $0$ or $1$. In addition to
$p_0$ and $p_2$ which are zero for the STI of the DP type, the infection
probabilities  $p_1$ and $p_4$ go to zero, and $p_3=P(1|011)=P(1|110)$
and $p_5=P(1|111)$ take the value $1$. We also note that $p_2$ is zero
in the non-spreading regime, as a single burst site with two laminar
neighbours is never observed in the non-spreading regime. Thus the
probability of infection of a laminar state by its neighbouring burst
state is zero in this regime, and no spreading of bursts can occur here. 
Cellular automata with probabilities which take values $0$ and $1$ alone
are called deterministic cellular automata (DCA), as given the state of
the system at a time $t$, its state at the time $t+1$ is
deterministically known. 
Thus the spatial intermittency in the non-spreading regime 
can be represented by a DCA, up to the coarse graining
defined earlier.    
Similar behaviour, either PCA (in the spreading regime) or DCA (in the non-spreading regime), is observed at other parameter points.  \\
A simple mean field approximation can be set up for the PCA
\cite{bagnoliatman}. Let $m_t$ and $m_{t+1}$ be the density of burst states in the lattice at
the $t^{th}$ and $t+1^{th}$ timestep. Using 
the CA rules defined above, the mean field evolution
equation  for the
density of bursts is given by $m_{t+1}= (2 p_1 + p_2) m_t(1-m_t)^2 
  + (2 p_3 +p_4) m_t^2 (1-m_t) + p_5 m_t^3$.

Approximating the values of Table \ref{pkdpsi} by 
$p_5=1$, and using $p_2=0$, 
the evolution equation reduces to
$m_{t+1}=  2 p_1~  m_t(1-m_t)^2 + 
 (2 p_3 +p_4)~ m_t^2 (1-m_t) + m_t^3$. This equation 
has
three fixed points
$m=0$, $ m=\frac{2 p_1-1}{1+2p_1-2p_3-p_4}=m^{\star}$, $m=1$.
\begin{figure}[!t]
\begin{center}                                                
\includegraphics[height=7.cm,width=10.5cm]{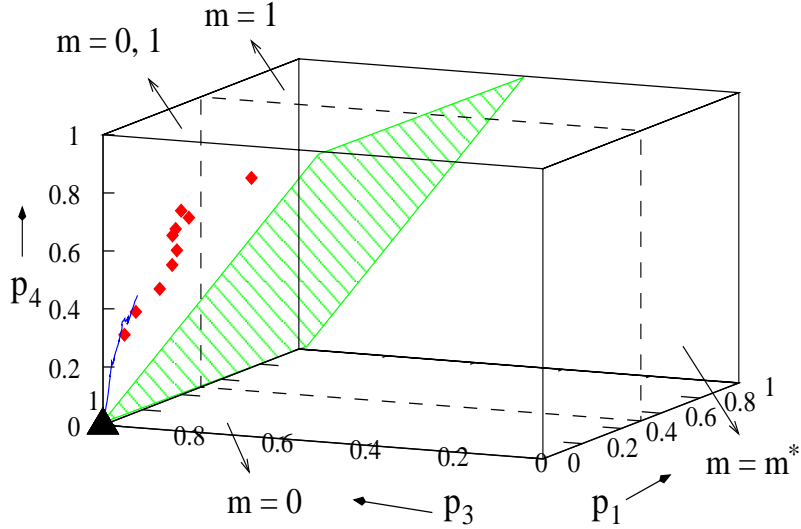}
\end{center}                                                  
\caption{(Color online) shows the $\{p_1,p_3,p_4\}$ probabilities of the DP points (marked with diamonds) and for points in the spreading regime at $\Omega=0.065$ (dotted line).   The triangle corresponds to the probabilities in the non-spreading regime.  \label{cubemf}}
\end{figure}             
The stability regions of these fixed points, as well as the 
co-existence region, where both the fixed
points $m=0$ and $m=1$ are stable,
 are shown in 
Figure \ref{cubemf}. 
The DP probabilities seen in  the spreading regime, as well as the PCA
probabilities seen at other points in the spreading regime,
lie in this co-existence region with the $2p_3 +
p_4=2$ plane  as a lower bound.  
However, all the probabilities associated with the deterministic cellular
automaton seen in the non-spreading regime (i.e. $ p_1=0,~p_3=1,~p_4=0
$) lie at the vertex  of the co-existence region in this cube (marked with a triangle in Fig. \ref{cubemf}).
As the parameters $\Omega$ and $\epsilon$ vary along 
curves which cross the infection line, the PCA probabilities seen in the 
spreading regime collapse to the DCA probabilities at the vertex of the
cube. 
Thus a PCA to DCA transition occurs at the infection line. 
Since the DP behaviour and the spatially intermittent behaviour along
the bifurcation boundaries  are  special cases
of spreading and non-spreading behaviour,
the transition from  the DP universality class of spatio-temporal intermittency to the  non-DP universality class
of spatial intermittency is reflected in the transition from
the PCA to the DCA 
\cite{chateca}.

\begin{figure}[!t]
\begin{center}                                                
\includegraphics[height=6.cm,width=8.5cm]{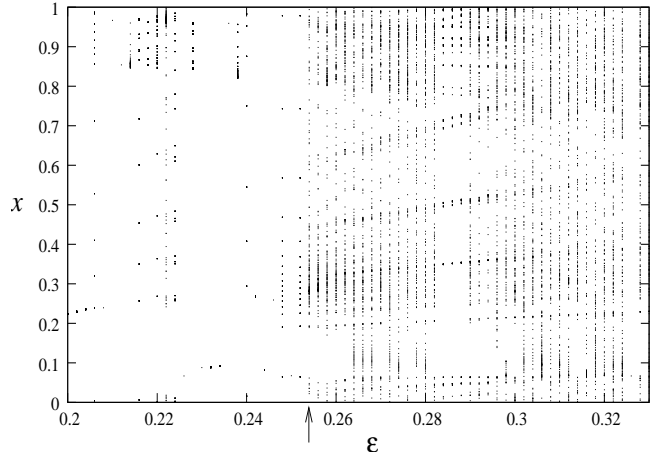}
\vspace{-.2in}                                                
\end{center}                                                  
\caption{shows the bifurcation diagram at $\Omega=0.065$ where $x$
values for a typical site $i$ have been plotted as a function of
$\epsilon$, an initial transient of $14,000$ iterates has been
discarded. \label{xbif}}
\end{figure}

The dynamical reason for this transition can be found by investigating 
the bifurcation diagram of the system (Fig. \ref{xbif}). 
The range of $\epsilon$
values on the vertical axis of Fig. \ref{pdsns} cut across the infection
line at $\epsilon=0.254$.  The bifurcation diagram clearly shows that an attractor widening
crisis \cite{greb} appears at this point. Similar behavior is seen for other sites. The spreading regime seen in
the phase diagram emerges exactly at the point at which the attractor
widens, with the non-spreading regime corresponding to the pre-widening
regime. This widening also identifies the point at which the equivalent
cellular automaton undergoes a PCA to DCA transition. In the pre-crisis
region, each site follows either a periodic or quasiperiodic trajectory 
and is not infected by the behaviour of its neighbours. Thus, its CA analogue
is deterministic as listed in Table \ref{pkdpsi}. In the postcrisis regime, 
each site is able to access the full $x$ range, as well as infect its
neigbours,and the bursting and spreading behaviour characteristic of the spreading regime is 
seen. This is reflected in the equivalent cellular automaton by a transition to PCA behavior (Table \ref{pkdpsi}).  
It is to be noted that the volume of the attractor in phase space
will be much larger post-crisis, as compared to the pre-crisis volume.
Further characterisation of the crisis is in progress.  

To conclude, 
the spatio-temporal intermittency of the directed percolation class and
the spatial intermittency of the non-directed percolation 
class, seen along the bifurcation boundaries of synchronised solutions,
are special cases of the spreading and non-spreading regimes seen off the
bifurcation boundaries. The two regimes are separated by the infection
line which intersects with the bifurcation boundary of the synchronised 
solutions at the point where the cross-over between the directed
percolation and non-directed percolation
behaviour takes place. Thus the behaviour seen in coupled sine circle map 
 lattice is organised around the locations of the bifurcation boundaries and
the synchronised solutions, and the infection line. The existence of two 
distinct universality
classes, 
in the phase diagram of the sine circle map lattice is a    
reflection of the  transition of the equivalent
cellular automaton from the probabilistic phase to the deterministic
phase and the concomitant suppression of the spreading or 
infectious modes. The dynamic origins of this transition lie in an
attractor-widening crisis. We believe this is the first time that such a
direct connection has been found between a dynamical phenomenon viz. a crisis 
in an extended system and the statistical properties of the extended
system viz. the exponents and universality classes. Similar directed
percolation to non-directed percolation  transitions have been seen in
other coupled map lattices, as well as in pair contact processes, solid on 
solid    
models  and models of non-equilibrium wetting \cite{Odor}. 
Our results may  have useful pointers for the analysis of other
systems, and thus contribute to the on-going 
debate on the identification of the universality classes of
spatiotemporal systems.    

ZJ thanks CSIR, India and NG thanks DST, India for partial support under the project SR/S2/HEP/10/2003.

\end{document}